\documentstyle[twocolumn,floats,aps]{revtex}

\input epsf         % defines \epsfbox and supporting macros
\epsfverbosetrue    % messages will show height and width

\begin{document}

%%%%%%%%%%%%%%%%%%%%%
\twocolumn[\hsize\textwidth\columnwidth\hsize\csname@twocolumnfalse%
\endcsname

\draft

\title{Spin Excitation in $d$-wave Superconductors :\\
A Fermi Liquid Picture}

\author{Khee-Kyun Voo, Hong-Yi Chen, and W. C. Wu}
\address{Department of Physics, National Taiwan Normal University,
Taipei 11650, Taiwan, R.O.C.}

\date{\today}

\maketitle

\begin{abstract}

A detailed study of the Inelastic Neutron Scattering (INS) spectra
of the high-$T_c$ cuprates based on the Fermi liquid (FL) picture
is given. We focus on the issue of the transformation between the
commensurate and incommensurate (IC) excitation driven by
frequency or $temperature$. For La$_{2-x}$Sr$_x$CuO$_4$ (LSCO),
the condition of small $\Delta(0)/v_F a$ (where $a$ is the lattice
constant, and henceforth will be set to 1) can simultaneously
reproduces the always existing IC peaks in the superconducting
(SC) and normal state, and the always fixed location at
temperature or frequency change. For YBa$_2$Cu$_3$O$_{6+x}$
(YBCO), a moderate $\Delta(0)/v_F a$ and proximity of the van Hove
singularity (vHS) at ${\overline M}=(0,\pi)$ to the Fermi level
can reproduce the frequency- and temperature-driven shifting IC
peaks in the SC state, and the vanishing of the IC peak in the
normal state. The commensurate peak is found to be more
appropriately described as a random phase approximation (RPA)
effect. We address the conditional peak shifting behavior to a
refined consideration on the nesting effect which is previously
overlook. As a result, both the data on LSCO and the recent data
on YBCO (on YBa$_2$Cu$_3$O$_{6.7}$ by Arai $et$ $al.$ and
YBa$_2$Cu$_3$O$_{6.85}$ by Bourges $et$ $al.$) can be reasonably
reconciled within a FL picture. We also point out that the
one-dimensional-like data by Mook $et$ $al.$ on a detwinned and
more underdoped sample YBa$_2$Cu$_3$O$_{6.6}$ could be due to a
gap anisotropy effect discussed by Rendell and Carbotte, and we
proceed to suggest a way of clarifying it.

\end{abstract}

\vskip 0.2 true in
%%%%%%%%%%%%%%%%%%%
]
%\narrowtext

\section{Introduction}
\label{sec:intro}

The aim of this paper is twofold. The first is to give a
comprehensive study of the essential properties of the BCS spin
susceptibility, both bare and RPA-corrected. The second, basing on
the simplest RPA theory we account for the present strongly
contrasting INS data on LSCO \cite{YLK98} and YBCO
\cite{ANE99,BSF00}.

%%%
Magnetic fluctuation in the high-$T_c$ cuprates has been long
believed to be very intimately related to their superconductivity
mechanism. Therefore various kinds of magnetic measurements such
as Nuclear Magnetic Resonance (NMR), Nuclear Quadrupole Resonance
(NQR), and INS have been carried out on the system, and aimed for
clarifying the connection between the two. Among those
measurements, INS experiments is the one which distinguishes
itself from others by its capability to measure the fluctuation
locally in momentum and frequency space and hence making itself an
indispensable and important tool. Especially recently it provides
a great deal of new observations, which has yet reached a
consensus.

%%%
Essentially there are two dominant features of the observed INS
spectra, the existence of commensurate peak at ${\bf
Q}_{AF}\equiv(\pi,\pi)$, and IC peak at ${\bf
Q}_\delta\equiv(\pi,\pi \pm \delta)$ and $(\pi\pm\delta, \pi)$.
The spectra in LSCO \cite{LAM99,YLK98,TAI97} is always IC and have
a incommensurability independent of $\omega$ and $T$ but increases
with hole-doping. On the other hand, YBCO \cite{BSF00,ANE99,DMD98}
and Bi$_2$Sr$_2$CaCu$_2$O$_{8+x}$ (BSCCO) \cite{FBS99,MDC98}
compounds exhibit both commensurate and IC peaks. Very recently a
comprehensive INS data on YBCO is obtained \cite{BSF00,ANE99}. In
the low-temperature SC state, it shows a commensurate peak at a
particular frequency $\omega_o(T=0)$. Departing from $\omega_o$
(either go below or above $\omega_o$), the peak is split up into
IC and the incommensurability is continuously increased. The IC
peak is found to appear only above some threshold frequency. In
the normal state, the excitation is always a broad commensurate or
weakly IC structure.  At rising temperature to $T_c$, the
incommensurability of the low frequency IC structure at low
temperature is continuously closed up to commensurate. Such
behavior is seen in an underdoped \cite{ANE99} and nearly
optimally doped YBCO \cite{BSF00} (while YBCO data on a broad
doping range are not available). These behaviors are in strong
contrast to LSCO, and therefore raise an important question as
whether the INS spectra in YBCO and LSCO are of the same origin.

%%%
There are also reports of the observation of the commensurate
resonance in the pseudo-gap phase of underdoped YBCO at the same
frequency as in the SC phase. But that remains controversial
\cite{BKR00}. At all doping levels, its resonance nature is only
well-established in the SC phase.

%%%
Perhaps the most straightforward and simplest interpretation of
the IC peak is the Fermi surface (FS) nesting effect (see e.g.,
Ref.~\cite{KSL00,Nor00,VW99,OP99,BL99} and references therein).
Many of the theoretical approaches assume that the system is
spatially homogeneous and ultimately end up at a FS, could it be a
FS of conventional electrons or a FS of some exotic fermion such
as the spinon. The spin excitation is constituted by those low
energy excitations nearby the Fermi surface. Such scenarios are
capable of producing frequency-shifted IC peaks
\cite{KSL00,Nor00,VW99}, and furthermore, the commensurate peak is
easily derived either from the bare or RPA susceptibility.

%%%
The Stripe ordering picture \cite{EK94} is an alternative which
predicts fixed IC peaks in spin and charge excitations that have
incommensurability directly proportional to, and depending only on
doping. In this picture the system is segregated into
one-dimensional hole-rich stripes in which the holes can freely
move, and electron-rich stripes in which electrons are ordered
antiferromagnetically. The electron filling determines the length
scale and thus the incommensurability. It is {believed} that the
static stripe at the 1/8-doping \cite{TSA95}, and the
2-to-1-coupled charge and spin excitations \cite{MD99} are
observed. That has boosted a lot of works on the ``dynamic''
stripe which also aims at the connection with the
superconductivity mechanism. A plain Stripe model predicts no
commensurate peak.

%%%
Therefore the Stripe picture seems to work well in LSCO, while the
FL picture works well in YBCO. This is intriguing if one adheres
to the philosophy that they all belong to the cuprate family and
therefore should lay in a single unified theory. Here we argue
that the discrepancies could be reconciled if the problem is
considered more carefully. The frequency and temperature driven
peak shifting present in YBCO but not in LSCO can be accounted for
by a flatter electronic dispersion near FS such that the
excitation can be scattered farther from it, while electronic
dispersion of LSCO is far more steep near FS. A refined
consideration of the nesting effect, which we have termed the
$dynamic$ $local$ $nesting$ effect is important when the condition
$\Delta/v_F\ll\pi$ is violated. In the regime of having shifting
IC peak in SC state, the IC peak is necessarily vanishes in the
normal state. The commensurate resonance in the SC state of YBCO
and BSCCO is addressed as a consequence of the experimentally
observed closeness of the vHS at ${\overline M}$ to the Fermi
level \cite{MQK95,GCA94,KSD94,DSK93,GCD93}. The effect of the band
singularity has been shown in many cases to be significant, such
as providing the vanishing isotope effect, transport anomalies,
high SC transition temperature, and so on. In short, we have
argued that the apparently contradictory INS data can have a
common basis in the FL regime.

%%%
We should also mention the data of Mook $et$ $al.$ \cite{MDD00} on
a detwinned and more underdoped YBCO, YBa$_2$Cu$_3$O$_{6.6}$. The
one-dimensional nature of the data where the IC peaks are found
only along one of the crystal axis, is claimed by the authors as a
strong evidence of stripe's existence. Nevertheless we point out
in Sec.~\ref{sec:aniso} that such behaviors can be due to an
anisotropy effect in the FL picture. The effect is prominent in
the frequency regime where the data in Ref.~\cite{MDD00} was
taken.

%%%
This study is also important in the aspect that it provides a
comprehensive and detailed survey of the major consequences from
the FL picture, and thus acts as a reference to see how much of
the experimentally observed behaviors are really deviating from
it. Comparisons with experiments are made where possible.

%%%
Our discussion will be focused on the evolution of the peak
intensity and location in the parameter space of $\omega$ and $T$.
We organize this paper into sections and a brief summary of the
findings of each sections or subsections is given at their ends
where necessary. The first several sections are focused on the
effect of dispersion $v_F$ on the IC excitation.
Sec.~\ref{sec:formalism} gives the formalism and explains how the
dispersion at Fermi level $v_F$ is modeled. Sec.~\ref{sec:survey}
gives the bare spectra in SC state, normal state, and at the
transition. Sec.~\ref{sec:dln} discusses the underlying machinery
of the formation and behaviors of the IC peaks. Sec.~\ref{sec:rpa}
studies the effect of inclusion of the Hubbard repulsion or AF
interaction into the system via RPA. An appropriate description of
LSCO is given. Sec.~\ref{sec:vhs} is devoted to the relation
between the INS peaks (commensurate and incommensurate) and
${\overline M}$-point vHS. An appropriate description of YBCO is
then given. Sec.~\ref{sec:aniso} is a discussion on the anisotropy
effect and the one-dimensional nature in INS spectrum. Finally,
Sec.~\ref{sec:conc} gives a discussion of the results and
conclusions.

\section{formalism}
\label{sec:formalism}

What we need in a FL description of the INS spectrum is the
existence of a well-defined FS and the dispersion nearby. All low
energy processes are constituted by the excitation at the vicinity
of the Fermi surface.

%%%
A INS spectrum is proportional to Im$\chi({\bf q},\omega)$ besides
some Bose-Einstein distribution factor due to the bosonic
character of the excitation. Here $\chi({\bf q},\omega)$ is the
spin susceptibility. The BCS bare spin susceptibility is given
below and followed by the RPA-corrected susceptibility.

%%%
The BCS bare spin susceptibility in a one-layer SC system (the
susceptibility in a normal system could be obtained by letting
$\Delta\rightarrow0$) is
\begin{eqnarray}
&&\chi_o({\bf q},\omega)=-{1\over 4}\sum_{\bf k}\left[\left[
1-{\xi_{\bf k}\xi_{\bf k+q}+\Delta_{\bf k}\Delta_{\bf k+q} \over
E_{\bf k}E_{\bf k+q}}\right]\right. \nonumber\\ &&\times
\left[\left.{1-f(E_{\bf k})-f(E_{\bf k+q})\over \omega-E_{\bf
k}-E_{\bf k+q}+i\Gamma}- {1-f(E_{\bf k})-f(E_{\bf k+q})\over
\omega+E_{\bf k}+E_{\bf k+q}+i\Gamma} \right]\right. \nonumber\\
&&-\left[\left.  1+{\xi_{\bf k}\xi_{\bf k+q}+ \Delta_{\bf
k}\Delta_{\bf k+q} \over E_{\bf k}E_{\bf k+q}}\right]\right.
\nonumber\\ &&\times \left[\left. {f(E_{\bf k})-f(E_{\bf
k+q})\over \omega-E_{\bf k}+E_{\bf k+q}+i\Gamma}- {f(E_{\bf
k})-f(E_{\bf k+q})\over \omega+E_{\bf k}-E_{\bf k+q}+i\Gamma}
\right]\right], \label{eq:chi0}
\end{eqnarray}
where {\bf q} and $\omega$ are respectively the momentum and
energy transfers. $f(E_{\bf k})$ is the Fermi function and $E_{\bf
k}=(\xi_{\bf k}^2+\Delta_{\bf k}^2)^{1/2}$ is the quasiparticle
spectrum with $\xi_{\bf k}$ and $\Delta_{\bf k}$ the band
dispersion and SC gap respectively. Henceforth we will write
$\chi_o' \equiv$ Re$\chi_o$, $\chi_o'' \equiv$ Im$\chi_o$, and
likewise for $\chi$ in Eq.~(\ref{eq:chi}).

%%%
We use a two-dimensional tight-binding electronic dispersion
\begin{equation}
\
\xi_{\bf k}=-2t(\cos k_{x}+\cos k_{y}) -4t'\cos k_{x}\cos
k_{y}-\mu, \label{eq:tbband}
\
\end{equation}
where $t$ and $t^\prime$ are the nearest-neighbor (NN) and
next-nearest-neighbor (NNN) hopping respectively. $\mu$ is the
chemical potential. In this paper only the $d_{x^2-y^2}$-gap is
treated (except in Sec.~\ref{sec:aniso}) since it is meant to
describe the high-$T_c$ cuprates. It is taken as $\Delta_{\bf
k}=\Delta(T)(\cos k_{x}-\cos k_{y})/2$. The width $\Gamma/t$ in
numerical integration is taken within $0.002\sim0.008$ with
slicing $2000\times2000$ or $4000\times4000$.

%%%
Equation~(\ref{eq:chi0}) describes two kinds of excitation, the
pair-breaking excitation that excites two quasiparticles from the
SC condensate and costs energy $E_{\bf k}+E_{\bf k+q}$, and the
thermal one-particle excitation that excites a quasiparticle from
${\bf k+q}$ to ${\bf k}$ which costs energy $E_{\bf k}-E_{\bf
k+q}$. The two-particle excitation vanishes in the normal state
while the one-particle excitation vanishes at zero-temperature SC
state.

%%%
Here we want to make several important remarks on the modeling [by
the simple dispersion Eq.~(\ref{eq:tbband})] of the dispersion
gradient at FS (Fermi velocity $v_F$) of a real system:

\begin{itemize}
\item[(i)] It is a fact that low energy physics is only relevant to the
part of the dispersion near the Fermi level. Therefore though the
real dispersion near the Fermi level can dramatically deviate from
the simple tight-binding band (sometimes even shows kink in the
dispersion \cite{BLK00}), for practical purposes we still can
model it by choosing an ${\bf effective~scale}~t$ that gives
similar ${\bf Fermi~velocity}$ $v_F$ at the Fermi level (since
$v_F=|d\varepsilon/d{\bf k}|_{\rm FS}$ $\propto t$). Thus in our
context, $\Delta/t$ is equivalent to $\Delta/v_F$, i.e. a greater
$\Delta/t$ could mean a smaller $v_F$ but does not necessarily
mean a greater $\Delta$. Obviously, we do not think ``$t$'' as the
bandwidth of the real dispersion, which is not of our concern.

\item[(ii)] On the ${\bf relative~size}$ of $\Delta/v_F$ in LSCO and YBCO:
Though there is a consensus that $\Delta_{\rm
YBCO}\sim2\Delta_{\rm LSCO}$, the $v_F$'s are very uncertain
\cite{note2,BLK00}. We therefore take $v_F$ as a phenomenological
input which is to be verified by future ARPES measurements, and
$v_F$ is thought as the dominant factor in the relative size of
$\Delta/v_F$.

\item[(iii)] Note also that since $v_F$ (or the effective $t$)
is uncertain, we will compare all energies (e.g. $\omega$, $T$,
etc) only in terms of $\Delta$.

\end{itemize}

%%%
The most important correction to the susceptibility should be the
existence of Coulomb or AF correlations between the
quasiparticles. Such correlations are believed to exist as a
residual interaction between the renormalized particles, and they
are conventionally treated by mean field decoupling as a
nontrivial step beyond bare theories. Then the susceptibility is
written into a RPA form as
\begin{equation}
\
\chi({\bf q},\omega)={{\chi_o({\bf k},\omega)}\over{1-{V_{\bf q
}}\chi_o({\bf k},\omega)}}, \label{eq:chi}
\
\end{equation}
where $V_{\bf q}$ can be the vertex of Hubbard repulsion $U$ or AF
interaction
\begin{equation}
\
J_{\bf q}=-{{|J|}\over2}(\cos{q_x}+\cos{q_y}). \label{eq:jq}
\
\end{equation}
$J_{\bf q}$ is different from $U$ essentially at being positive
near ${\bf Q}_{AF}$ and changing its sign to negative near
$(0,0)$. The RPA-type correction is important because it is
perhaps the only way to treat correlation effect analytically.

\section{excitation at bare level\\
----- survey of basic properties}
\label{sec:survey}

We have given in this section a survey of the essential properties
of the bare excitation spectrum. Their discussion will be given in
Sec.~\ref{sec:dln}.

\begin{center}
{\bf A. Superconducting State}
\end{center}

The SC state is described by $\Delta(T)\neq0$, which results in a
quasiparticle excitation gap and a SC coherence factor.

%%%
\begin{figure}[t]
\begin{center}
\vspace{-0.0cm} \leavevmode\epsfxsize=3.2in \epsfbox{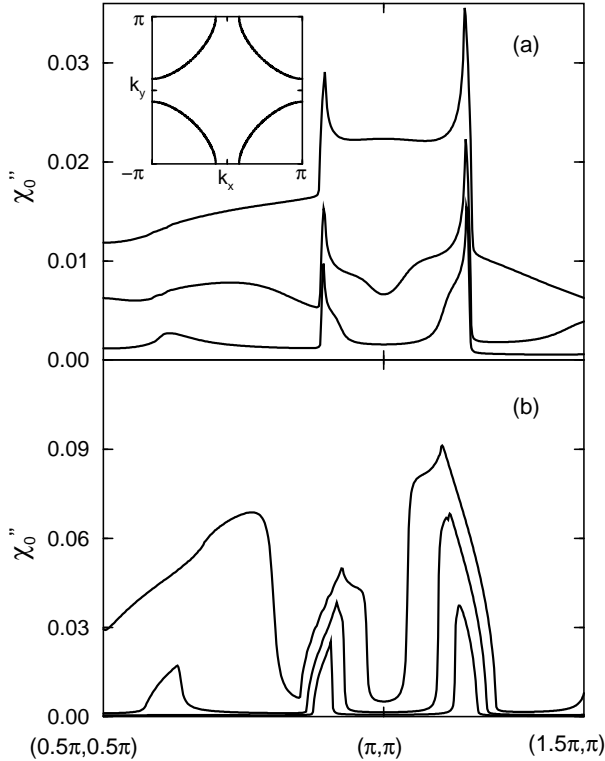}
\vspace{-0.0cm}
\end{center}
\caption{Frequency evolution of the SC state $\chi_o''({\bf q},
\omega)$ with different $\Delta/t$ is shown. (a) $\Delta/t=0.03$,
$\omega/\Delta=1.0,1.5$, and $2.0$ (from bottom to top); (b)
$\Delta/t=0.30$, $\omega/\Delta=0.6,1.0$, and $1.4$ (from bottom
to top). In (b), the broad peak shifts at changing frequency, and
some stray structure away from the IC peak is present. Including
the RPA-correction [see Fig.~\protect\ref{rpa3}(d)] will make the
IC peaks and shifting effect more conspicuous. We particularly
note that since we mean to describe two $different$ systems in (a)
and (b), the ``$\Delta$'' or ``t'' in (a) and (b) are unrelated
quantities. For both panels, $T=0$, the dispersion is $t=1$,
$t'=-0.25$, and $\mu=-0.65$. The FS is shown in the inset and the
plots are along ${\bf q}= (0.5\pi,0.5\pi)-(\pi,\pi)-(1.5\pi,\pi)$.}
\label{sura1} \vspace{-0.3cm}
\end{figure}
%%%

Figure~\ref{sura1} shows the frequency-evolution of the spectra
for two typical cases of $\Delta/t$. We have ignored the
well-discussed quasielastic node-to-node IC peaks \cite{Lu92} at
$(\pi + \delta_o,\pi \pm \delta_o)$ and $(\pi - \delta_o , \pi \pm
\delta_o)$, and focused on the nesting IC peaks at $(\pi,\pi \pm
\delta)$ and $(\pi \pm \delta, \pi)$. In general, as $\omega$ is
increased, the intensity and widths of the peaks are increased,
and spectrum at ${\bf Q}_{AF}$ is filled in at some frequency near
$2\Delta$ which depends only on band geometry. When $\Delta/t$ is
large, the fill-in is a stepwise jump of $\chi_o ''({\bf
Q}_{AF},\omega)$. This sharp increase of $\chi_o ''({\bf
Q}_{AF},\omega)$ has an important relation to the resonance at
${\bf Q}_{AF}$ and will be further discussed in
Sec.~\ref{sec:vhs}. The nesting IC peaks are seen to be sharpest
at some intermediate frequencies before the filling in.

%%%
In contrast to the sharp, well-preserved and not-shifted peaks in
the $\Delta/t\ll1$ case [see Fig.~\ref{sura1}(a)], the larger
$\Delta/t$ calculation [see Fig.~\ref{sura1}(b)] shows broader and
biased peaks that clump into ${\bf Q}_{AF}$ with increasing
frequency. Such converging of the IC peak cluster was observed
experimentally in YBCO and theoretically reproduced by us
\cite{VW99} and other groups \cite{Abr00,Nor00,KSL00}. Here we
proceed to give its full explanation in Sec.~\ref{sec:dln}. If the
saddle vHS at ${\bf k}=(0,\pi)$ is far from the Fermi level, the
IC structure will be destroyed after the peaks clump in at some
$\omega/\Delta\lesssim2$.

%%%
Some stray structures around the IC peaks always exist especially
in the case of larger $\Delta/t$ and $\omega/\Delta$. When the RPA
correction is introduced, they will become relatively weaker.

%%%
The main point here is that at $\omega\sim\Delta$, a system with
moderate $\Delta/v_F$ (modeled by a moderate $\Delta/t$) has its
nesting peak broad and driven by $\omega$ to shift. We will
discuss in Sec.~\ref{sec:dln} that the shift direction actually
depends on the orientation of the curvature of the FS near the gap
nodes.

\begin{center}
{\bf B. Normal State}
\end{center}

The normal state is described by $\Delta(T)=0$, which leaves the
bandwidth or the Fermi velocity as the only energy scale.
Eq.~(\ref{eq:chi0}) is then reduced to the Lindhard function
\begin{eqnarray}
&&\chi_o({\bf q},\omega)=\sum_{\bf k} { {f(\xi_{\bf k})-f(\xi_{\bf
k+q})} \over { \omega-\xi_{\bf k}+\xi_{\bf k+q}+i\Gamma } }
\label{eq:lindhard}
\end{eqnarray}
via $E_{\bf k}\rightarrow|\xi_{\bf k}|$, $|\xi_{\bf k}|=\xi_{\bf
k}$ and $-\xi_{\bf k}$ when $\xi_{\bf k}>0$ and $\xi_{\bf k}<0$
respectively. The relation $f(-\xi_{\bf k})=1-f(\xi_{\bf k})$ is
also used.

%%%
\begin{figure}[t]
\begin{center}
\vspace{-0.0cm} \leavevmode\epsfxsize=3.2in \epsfbox{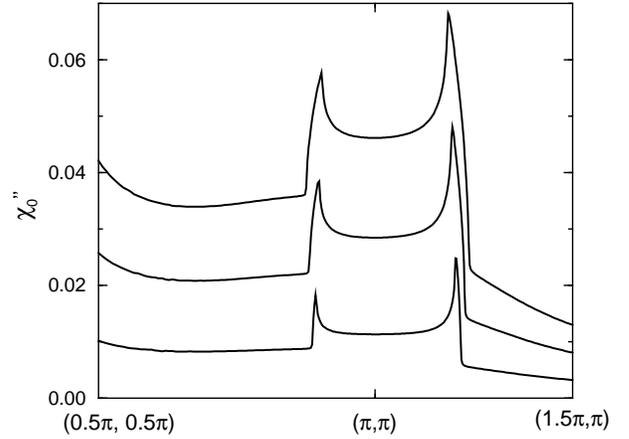}
\end{center}
\caption{Normal state (i.e. $\Delta=0$) $\chi_o''({\bf q},
\omega)$ at $\omega/t=0.04, 0.10$, and $0.16$ (from bottom to top)
is shown along ${\bf q}= (0.5\pi,0.5\pi)-(\pi,\pi)-(1.5\pi,\pi)$.
Such IC peaks exist only in the regime $\omega/t\ll1$. Temperature
$T=0$ and the dispersion is $t=1$, $t'=-0.25$, and $\mu=-0.65$.}
\label{surb1} \vspace{-0.3cm}
\end{figure}
%%%

Figure~\ref{surb1} shows some typical normal state spectra. The IC
peaks are found to exist only at the regime $\omega/t\ll1$. Such
destruction of the normal state peaks by frequency qualitatively
agrees with the experimental fact \cite{MAM92}. In that regime,
increasing frequency broadens the peak and enhances the intensity,
but do not shift the peaks. Similar to the discussion in previous
subsection, $\omega/t\ll1$ in our context could simply mean a
steep dispersion at FS (i.e. large $t$) and does not necessarily
mean a small $\omega$.

\begin{center}
{\bf C. Transition between\\ Superconducting and Normal States}
\end{center}

\begin{figure}[t]
\begin{center}
\vspace{-0.0cm} \leavevmode\epsfxsize=3.2in \epsfbox{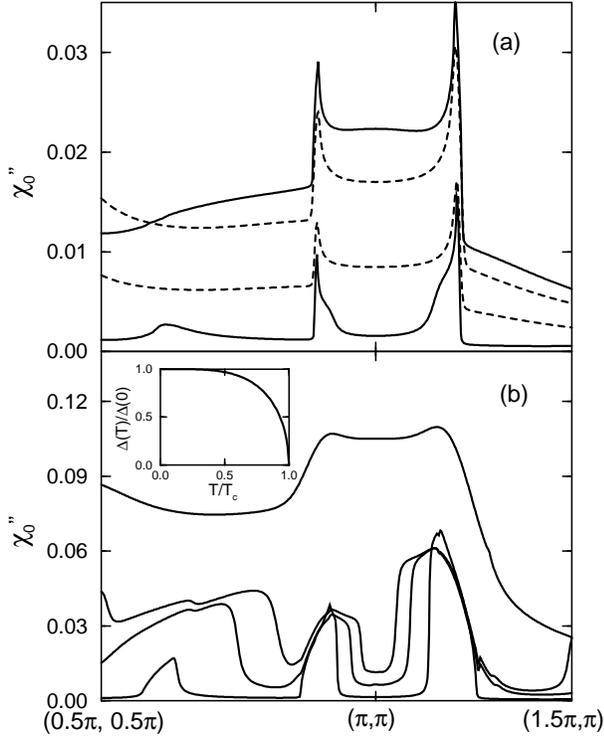}
\vspace{-0.0cm}
\end{center}
\caption{Bare spectra $\chi_o''({\bf q},\omega)$ with different
$\Delta(0)/t$ and $\omega/\Delta(0)$ at varying temperature. (a)
$\Delta(0)/t=0.03$: the solid lines show the $T=0$ spectra at
$\omega/\Delta(0)=1.0$ and $2.0$ (from bottom to top). The dashed
lines has likewise frequencies but $T=T_c=0.25\Delta(0)$. (b)
$\Delta(0)/t=0.30$ and $\omega/\Delta(0)=1.0$:
$T/T_c=0,0.80,0.85$, and $1.0$ (from bottom to top).
$T_c=0.25\Delta(0)$ and the assumed $T$-dependence of $\Delta(T)$
is shown in the inset. Heating converges the IC peak cluster in
(b) to a commensurate structure. Note also that neither the
``$\Delta$'' or ``t'' in (a) is related to that in (b). The
dispersion used in both panels is $t=1$, $t'=-0.25$, and
$\mu=-0.65$. The plots are along ${\bf
q}=(0.5\pi,0.5\pi)-(\pi,\pi)-(1.5\pi,\pi)$.} \label{surc1}
 \vspace{-0.3cm}
 \end{figure}

In the transition from SC to normal state, the gap is continuously
diminished and hence the coherence factor. The restriction on
transition in phase space is also relaxed. We will take an
empirical relation $\Delta(T)=\Delta(0)[1-(T/T_c)^4]^{1/2}$ [see
inset in Fig.~\ref{surc1}] at $T<T_c$ in our discussion.

%%%
%%%

If the IC peaks exist in the normal state, at entering the SC
state [see Fig.~\ref{surc1}(a)] it will be enhanced when
$\omega/\Delta(0)>1$ and suppressed if $\omega/\Delta(0)\sim1$,
and it will also be sharpened for $\omega/\Delta(0)<2$ since
spectral weight at ${\bf Q}_{AF}$ will be gapped out.

%%%
Figure~\ref{surc1}(b) shows a remarkable case of transition across
$T_c$. For a system which was shown previously to have
frequency-shifted IC peaks in the SC state [see
Fig.~\ref{sura1}(b)], increasing temperature broadens the peaks
and shifts them towards ${\bf Q}_{AF}$. The peaks merge into a
broad commensurate (or weakly IC structure) structure in the
normal state. The change is drastic at $T \lesssim T_c$ where
$\Delta(T)$ is rapidly changing.

%%%
The essential point here is the following. Given $\Delta$ of a
system, when $v_F$ (model by $t$) is such that $\Delta/v_F$
(modeled by $\Delta/t$) is large enough to provide
frequency-shifted IC peaks, the peaks will also be
temperature-shifted and vanish in the normal state (since the
condition $\omega/t\ll1$ for the normal state peak to exist is
necessarily violated at $\omega\sim\Delta$).

\section{excitation at bare level\\
----- the dynamic local nesting}
\label{sec:dln}

The IC peaks are originated from the $umklapp$ and inversion
symmetry of the dispersion. This section discusses the properties
of the bare excitation mentioned in Sec.~\ref{sec:survey}. Both
the $\omega$- and $T$-driven peak shifting effects are essentially
due to the dynamic and local nature of the nesting effect.

We start to illustrate the mechanism by a zero temperature SC
system where the bare spectrum is given by
\begin{eqnarray}
&&\chi_o''({\bf q},\omega)={\pi\over 4}\sum_{\bf k}\left[\left.
1-{\xi_{\bf k}\xi_{\bf k+q}+\Delta_{\bf k}\Delta_{\bf k+q} \over
E_{\bf k}E_{\bf k+q}}\right]\right. \nonumber\\ &&\times
\left[\left.{\delta (\omega-E_{\bf k}-E_{\bf k+q})}- {\delta
(\omega+E_{\bf k}+E_{\bf k+q})} \right]\right. . \nonumber\\
\label{eq:chi011t0}
\end{eqnarray}
The integrand consists of a coherence factor which reflects the
non-time-reversal invariance nature of the magnetic measurement,
and a $\delta$-function that imposes the energy conservation rule.

%%%
The role of the $\delta$-function should be first discussed since
it is the most dominant. It limits the contributing region of the
phase space by the energy conservation $E_{\bf k}+E_{\bf
k+q}=\omega$. At $\omega\sim0$, the only possibility is $E_{\bf
k}$ and $E_{\bf k+q}$ both $\sim0$. Therefore {\bf q} can only be
vectors connecting the gap nodes, which are ${\bf
Q}_{\delta_o}\equiv(\pi + \delta_o,\pi \pm \delta_o)$ and $(\pi -
\delta_o , \pi \pm \delta_o)$, with
$\delta_o=2\sin^{-1}\sqrt{\left[t-\sqrt{t^2-t'\mu}\right]/
\left[-2t'\right]}$.

%%%
At $\omega>0$, to satisfy $E_{\bf k}+E_{\bf k+q}=\omega$, ${\bf
q}$ can be any momentum that connects points on the contours
$E_{\bf k}=\omega_1$ and $E_{\bf {k+q}}=\omega_2$, if
$\omega_1+\omega_2=\omega$. These are strips of area stretched out
from the gap node along the FS. For most {\bf q}, it only
constitutes a featureless and ``incoherent'' background in the
{\bf q}-space. To constitute a salient structure out of the
background one needs to have extra factors that can stack extra
weight onto the background. A situation that can further provide
some weight is when $\omega_1=\omega_2=\omega/2$ and all the two
frequency dependent contours become $E_{\bf k}=E_{\bf
k+q}=\omega/2$. It spans the whole FS when
$\omega\rightarrow2\Delta$. Now ${\bf q}$ can ``nest" two contours
locally by passing through points ${\bf k}_o=(m\pi,n\pi)$ (where
$m$ and $n$ are integers) in the BZ. It occurs because the
$\omega$-dependent contour is always locally parallel at ${\bf
k}_o+{\bf k}$ and ${\bf k}_o-{\bf k}$ due to the dispersion
$E_{\bf k}$ has the ${\bf k}\leftrightarrow{\bf -k}$ inversion and
$umklapp$ symmetry in the extended BZ. This nesting is thus
$dynamic$ and $local$.

%%%
\begin{figure}[t]
\begin{center}
\vspace{-0.0cm} \leavevmode\epsfxsize=3.3in \epsfbox{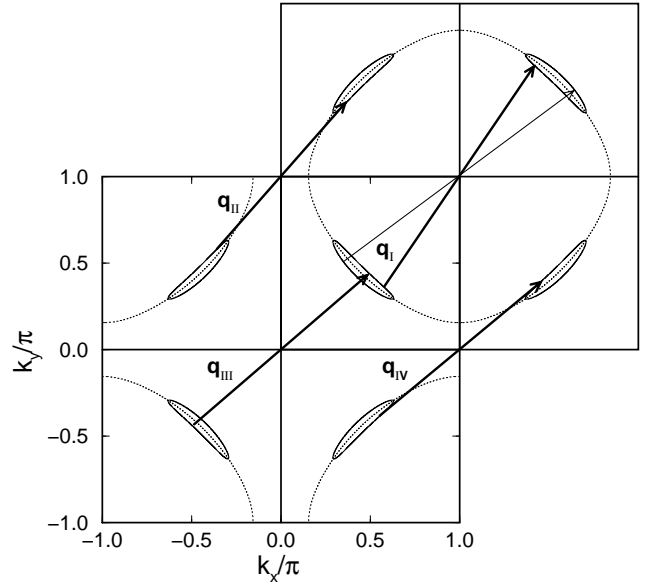}
\vspace{-0.5cm}
\end{center}
\caption{Energy contour $E_{\bf k}=\omega/2$ in the extended BZ
for $\omega=1.0\Delta(0)$. Some arbitrary local-nesting vectors
${\bf q}_{I}$, ${\bf q}_{II}$, ${\bf q}_{III}$, and ${\bf q}_{IV}$
are shown. The thick vector between above-FS contour piece have
better nesting due to the smaller curvature. The FS is shown as
dotted line, and the dispersion is $t=1$, $t'=-0.25$, and
$\mu=-0.65$. The gap here is chosen as $\Delta/t=0.50$.}
\label{dln1}
\end{figure}
%%%
As an example, we consider $\chi_o''({\bf q},\omega)$ at ${\bf
q}\in[0\sim2\pi,0\sim2\pi]$ and the integration over {\bf k} in
evaluating $\chi_o''$ is run over
$(k_x,k_y)\in[-\pi\sim\pi,-\pi\sim\pi]$. Fig.~\ref{dln1}
illustrates the local-nesting at a frequency $\omega<2\Delta$. In
the integration over the first quadrant
$(k_x,k_y)\in[0\sim\pi,0\sim\pi]$, those local nesting vectors are
shown as ${\bf q}_{I}$. Likewise, ${\bf q}_{II}$ for the second,
${\bf q}_{III}$ for the third, and ${\bf q}_{IV}$ for the fourth
quadrant.

%%%
\begin{figure}[t]
\begin{center}
\vspace{-0.1cm} \leavevmode\epsfxsize=3.4in \epsfbox{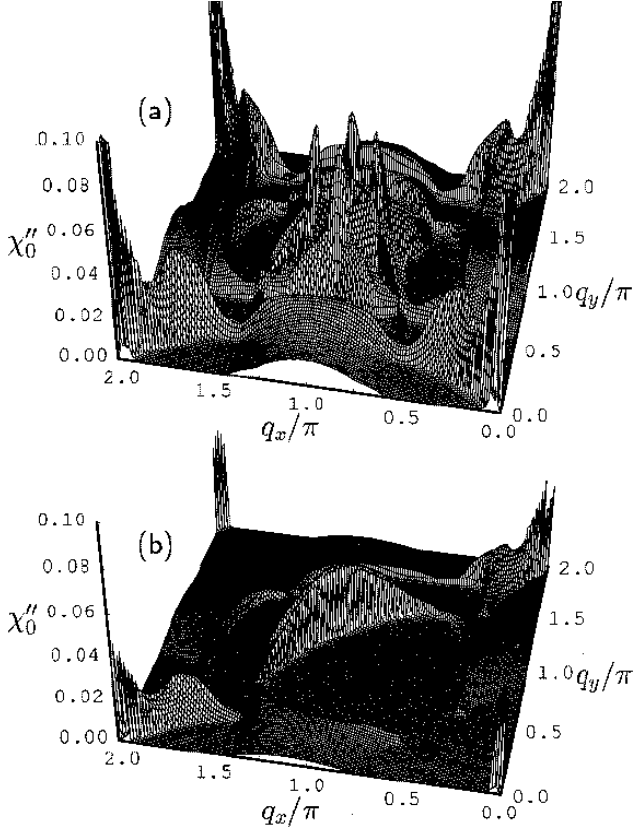}
%\vspace{-2.3cm}
\end{center}
\caption {(a) A bare spectrum $\chi_o''({\bf q},\omega)$ in the
full {\bf q}-space at $\omega=1.0\Delta(0)$. (b) a single branch
of ``ridge'', obtained when the integration over {\bf k }in
Eq.~(\protect\ref{eq:chi0}) is done only over the quadrant
$(k_x,k_y)\in[0\sim\pi,0\sim\pi]$. Four of these 90-degree rotated
branches superpose to give $\chi_o''({\bf q},\omega)$ in (a).
Refer the dispersion parameters to Fig.~\protect\ref{sura1}(b).}
\label{dln2}  \vspace{-0.3cm}
\end{figure}
%%%
Figure~\ref{dln2}(a) shows the consequence of such local-nesting
effect. The formation of the structure can be illustrated by an
illuminating calculation, where the integration is done only over
the first quadrant $(k_x,k_y)\in[0\sim\pi,0\sim\pi]$. Then only
one branch of the ridge-like structure is obtained [see
Fig.~\ref{dln2}(b)] and its orientation is the same as the contour
in quadrant $(k_x,k_y)\in[\pi\sim2\pi,\pi\sim2\pi]$. The ends of
the ridge are determined by $\Delta_{\bf k}=\omega/2$. The locus
of the ridge in {\bf q}-space resembles the locus of the contour
(which is approximately the FS) in {\bf k}-space but have twice
the extension since the nesting vectors are twice the vectors
describing the locus of the FS in {\bf k}-space [see
Fig.~\ref{dln1}].

%%%
An IC peak is a result of overlapping of two ridges at the
symmetrical points ${\bf Q}_\delta=(\pi,\pi\pm\delta)$ and
$(\pi\pm\delta,\pi)$, where $\delta=2\sin^{-1}[-\mu/(2t)]$, when
$\omega>\omega_{\delta}\equiv\Delta(T)[-\mu/(2t)]$. For those FS
of open topology [see Fig.~\ref{dln1}], the ridges overlap also at
${\bf Q}_{\delta'}=(\pi+\delta',\pi\pm\delta')$ and
$(\pi-\delta',\pi\pm\delta')$ when
$\omega>\omega_{\delta'}\equiv\Delta(T)\sqrt{\mu/t'}$, where
$\delta'=2\sin^{-1}\sqrt{\mu/(4t')}$. The above expressions for
$\delta_o, \delta,$ and $\delta'$ are estimations since the ridges
have finite thickness.

%%%
Frequency $\omega_{\delta'}\equiv\Delta(T)\sqrt{\mu/t'}$ is also
the onset frequency of $\chi_o''({\bf Q}_{AF},\omega)$, since
${\bf Q}_{AF}$ also connects equivalent points of ${\bf
Q}_{\delta'}/2$ (also called the ``hot spots'').

%%%
Due to the bending of the ridges [see Fig.~\ref{dln2}], one will
miss the nodal excitation point ${\bf Q}_{\delta_o}$ if one scan
two nesting IC peaks at ${\bf Q}_\delta$ in their nearest distance
direction. Therefore $\chi_o''$ will be seen as always gapped at
any {\bf q}. Since this should be more prominent when $\Delta/t$
is small (which is likely the case of LSCO), this may explain the
observation of a momentum independent spin gap by Lake $et$ $al$.
on LSCO \cite{LAM99}.

%%%
A very peculiar feature which has been overlooked previously is
the peak shifting behavior, brought by the difference of curvature
of the contour pieces below and above the FS near the gap nodes.
This occurs when the nodal FS segment has nonzero curvature and
the contour $E_{\bf k}=\omega/2$ is appreciably opened up from it.
In the geometry of FS shown in Fig.~\ref{dln1}, the above-FS
contour has a smaller curvature and can be nested more
effectively. That causes the ridges to have biased cross-sections,
which in turn causes the nesting peaks to have their apex biased
to some direction [see Fig.~\ref{sura1}(b)]. The opening up of the
contour with $\omega$ increases the widths of the ridges and
peaks, and therefore shifts the apex of the broad peaks [see
Fig.~\ref{sura1}(b)]. The broad and shifting behavior of the bare
excitation peak are thus necessarily coexist.

%%%
Finite temperature in SC states drives the gap and thus changes
the extensions of ridges both along and away from the FS. When $T$
broadens the ridges away from FS, it simultaneously shifts the
peaks apex like increasing $\omega$. Besides, $T$ always causes
some smearing in all cases, the SC or the normal state.

%%%
When the dispersion at FS (i.e. $v_F$ or $t$) is steep or the gap
($\Delta$) is small, for $\omega\sim\Delta$, the contours are
confined on the FS because its transverse extension in phase space
$\delta{\bf k}$ $\sim$ $\Delta/|d\varepsilon/d{\bf k}|_{\rm
FS}=\Delta/v_F\ll\pi$ (this is equivalent to the condition
$\Delta/t\ll1$). The ridges and IC peaks formed are slim and no
shift of peak is possible. For the peak to shift, a finite
curvature of the FS is also necessary to provide the difference of
curvature between the above- and below-FS contours.

%%%
The coherence factor has values between $0$ and $2$, and that can
provide modification of the intensity as pointed out in the
literature. The condition $\xi_{\bf k}\rightarrow0$ at the FS
tends to maximize the spin coherence factor to $2$ at $\Delta_{\bf
k}\Delta_{\bf k+q}<0$, and minimize it to $0$ at $\Delta_{\bf
k}\Delta_{\bf k+q}>0$. As our nesting vectors has $\Delta_{\bf
k}\Delta_{\bf k+q}>0$, the ends of ridges are smoothly suppressed
by the cutoff $\Delta_{\bf k}=\omega/2$. In the case of larger
$\Delta/t$, the condition $\xi_{\bf k}\rightarrow0$ is not well
satisfied at finite $\omega/\Delta$ since the contour is opened up
appreciably from the FS. This weakens the suppression effect
before cutoff, and the ridges will overlap at more sizable values
and gives a higher contrast of the nesting IC peaks. It also
lowers the threshold $\omega/\Delta$ for the emergence of the
peaks.

%%%
The normal state is the lift of the energy restriction by the gap,
and the disappearance of coherence factor. The ridges now always
span the whole {\bf q}-space and the nesting peaks always exist in
principle. But due to replacing a stronger energy constraint (due
to two-particle excitation) by a looser constraint (due to
one-particle excitation), the transition now floods over regions
far away from the FS and makes the widths of ridges dispersed and
the nesting peaks less distinguishable. Especially now the
transition near ${\bf Q}_{AF}$ is filled high for hole-like FS.

%%%
Albeit the underlying cause of formation of the ridges and nesting
peaks ($umklapp$ and {\bf k}-inversion symmetry) is irrelevant to
the existence or not of the SC gap, there are still some upper
limits on $\omega/\Delta$ or $\omega/t$ for the IC peaks to exist.
Since increasing $\omega/t$ broadens the ridges, the ridge and
peak structure will be always destroyed at very high $\omega/t$.
The same is true for very high $T/t$ where the excitation is
deconfined from the FS. It is easier for the peak to survive in
the SC state because the near ${\bf Q}_{AF}$ transition is kept
away at $\omega/\Delta<2$ [see Fig.~\ref{surc1}(b)].

%%%
To summarize, at low $\omega/t$ and $T/t$, the nesting IC peaks
should always exist in bands with $umklapp$ and inversion (${\bf
k}\leftrightarrow{\bf -k}$) symmetry. In the SC state, it can be
driven by $\omega$ or $T$ to shift when $\Delta/t$ is appreciable
and the nodal FS is curved.

\section{RPA-corrected excitation}
\label{sec:rpa}

In this section we will discuss the effect of RPA correction on
the frequency-driven shifting effect of the IC peaks. We start
with a brief discussion of the essential features of the RPA
spectrum.

%%%
The RPA spectrum is given by
\begin{equation}
\
\chi''({\bf q},\omega)={ {\chi_o''({\bf q},\omega)} \over
{[1-V_{\bf q}\chi_o'({\bf q},\omega)]^2+ [V_{\bf q}\chi_o''({\bf
q},\omega)]^2} }. \label{eq:chi11}
\
\end{equation}
It can be viewed as a scaling of $\chi_0''$ by the Stoner factor
$[(1-V_{\bf q}\chi_o')^2+ (V_{\bf q}\chi_o'')^2]$, which is
greater or equal to zero.

%%%
\begin{figure}[t]
\begin{center}
\vspace{-0.0cm} \leavevmode\epsfxsize=3.2in \epsfbox{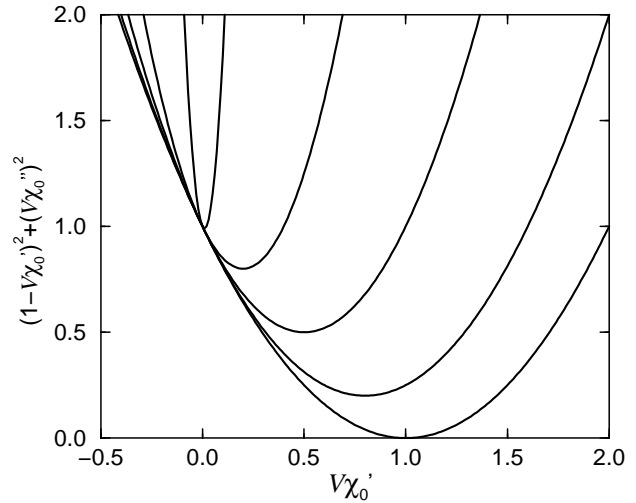}
\vspace{-0.0cm}
\end{center}
\caption {The Stoner factor $[(1-V\chi_o')^2+ (V\chi_o'')^2]$ is
plotted against $V\chi_o'$ for $\chi_o''/\chi_o'=0, 0.5, 1.0,
2.0$, and $10$ (from bottom to top).}
\label{rpa1}  \vspace{-0.3cm}
\end{figure}

%%%
Figure~\ref{rpa1} shows that Stoner factor $s=[(1-V\chi_o')^2+
(V\chi_o'')^2]$ versus $V\chi_o'$ for various $\chi_o''/\chi_o'$.
It summarizes all features of the RPA correction in a lucid way.
For instance, at the regime of small $\chi_o''/\chi_o'$, RPA
spectra with $V<1/\chi_o'$ is just a scaling (i.e. $s\neq0$, could
be magnification or suppression) of the bare spectra; while RPA
spectra with $V\simeq1/\chi_o'$ are resonances (i.e. $s\sim0$)
with damping $\chi_o''$. For large $\chi_o''/\chi_o'$, $s$ is
greater than 1 in most regions, which means that for most choices
of $V$ the RPA correction has only suppression effect. This can
happen in systems with large $\chi_o''$ due to band singularity.

%%%
\begin{figure}[t]
\begin{center}
\vspace{-0.0cm}
\leavevmode\epsfxsize=3.4in \epsfbox{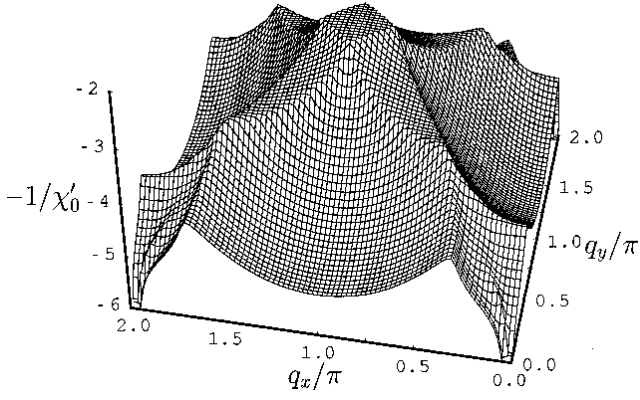}
%\vspace{-4.0cm}
\end{center}
\caption {A typical SC state $-1/\chi_o'({\bf q},\omega=0)$ in
{\bf q}-space. The no-instability bound of the interaction
strength is $V_{\bf q}<1/\chi_o'({\bf q},\omega=0)$. For $V_{\bf
q}=J_{\bf q}$, $|J|<1/\chi_o'({\bf q}\sim{\bf
Q}_{AF},\omega=0)\simeq2.2 t$ in this case. The plot has $T=0$,
$\Delta/t=0.03$, and band dispersion $t=1$, $t'=-0.25$, and
$\mu=-0.65$.}
\label{rpa2} \vspace{-0.3cm}
\end{figure}
%%%
In reality, the RPA resonance could be unlikely to occur in
systems with well-behaved properties near the Fermi level by the
following reason. For any stable system, $V_{\bf q}$ should be
sufficiently away from the instability boundary $V_{\bf
q}=1/\chi_o'({\bf q},\omega=0)$ [see Fig.~\ref{rpa2} for an
example]. When there is no singular sources such as the van Hove
singularity near the Fermi level, $\chi''_o({\bf q},\omega)$
changes smoothly with $\omega$ and $\chi_o'({\bf q},\omega)$ is
inert to change of $\omega$ (due to the Kramers-Kronig relation).
Theoretically a fine tuning of the interaction strength is then
required to get a resonance. In such systems, absence of
instability at zero $\omega$ is likely to imply absence of
resonance at nonzero $\omega$.

%%%
\begin{figure}
\begin{center}
\vspace{-1.0cm}
\leavevmode\epsfxsize=3.4in \epsfbox{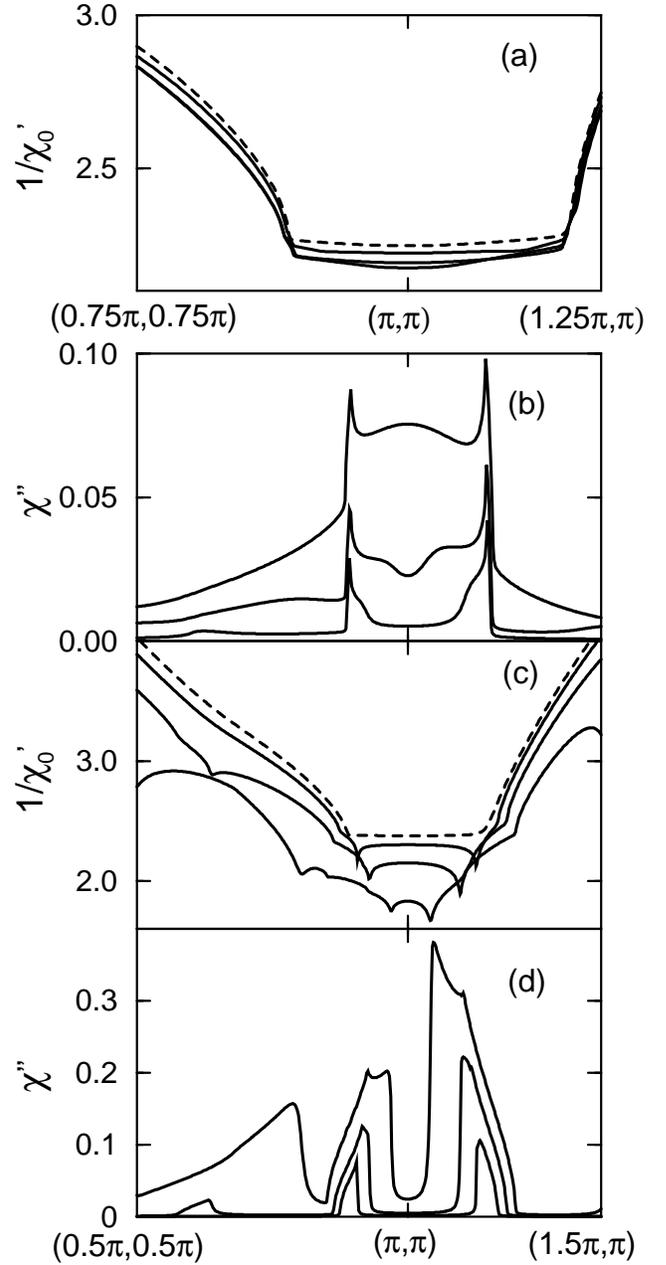}
\vspace{0.3cm}
\end{center}
\caption{RPA effect (with no resonance) on two systems with small
and moderate $\Delta/t$. All graphs except (a) are plotted along
${\bf q}=(0.5\pi,0.5\pi)-(\pi,\pi)-(1.5\pi, \pi)$. For the case of
$\Delta/t=0.03$ and $|J|/t=1.0$: (a) $1/\chi_o'({\bf q},\omega)$
is shown [along ${\bf
q}=(0.75\pi,0.75\pi)-(\pi,\pi)-(1.25\pi,\pi)$] at
$\omega/\Delta=0$ (dashed line), $1.0,2.0$ and $1.5$ (solid lines
at ${\bf Q}_{AF}$ from top to bottom); Correspondingly, (b)
$\chi''({\bf q},\omega)$ is shown at $\omega/\Delta=1.0,1.5$ and
$2.0$ (from bottom to top). For the case of $\Delta/t=0.30$ and
$|J|/t=1.0$: (c) $1/\chi_o'({\bf q},\omega)$ is shown at
$\omega/\Delta=0$ (dashed line), $0.60,1.0$ and $1.4$ (solid lines
from top to bottom); Correspondingly, (d) $\chi''({\bf q},\omega)$
is shown at $\omega/\Delta=0.6,1.0$ and $1.4$ (from bottom to
top). Note the sharp minima of $1/\chi_o'$ in (c) and its
discussion in the text. Temperature $T=0$ and the corresponding
bare spectra were shown previously in Fig.~\protect\ref{sura1}.}
\label{rpa3} \vspace{-0.3cm}
\end{figure}

%%%
Figure \ref{rpa3} illustrates the effect of RPA correction on the
peak shifting behavior. We have considered antiferromagnetic
interactions $J_{\bf q}$ which do not cause resonance (weak
coupling). In the case of moderate $\Delta/t$, sharp local minima
of $1/\chi_o'$ [Fig. \ref{rpa3}(c)] appear at the sharp edges of
the bare IC peaks due to the Kramers-Kronig relation. This is
nontrivial as it results in more spectacular shifting of the IC
peaks [compare Fig. \ref{rpa3}(d) and Fig. \ref{sura1}(b)]. On the
other hand, in the case of small $\Delta/v_F$, the fixed
incommensuration behavior is maintained [Fig. \ref{rpa3}(b)] due
to the ``quiescent'' $1/\chi_o'$ [Fig. \ref{rpa3}(a)]. We conclude
that the weak coupling RPA correction do not alter the original
behaviors of the IC peaks, and the LSCO data can be described by
small $\Delta/v_F$ and weak coupling RPA correction.

\section{excitation spectrum and\\
${\overline M}$-point van Hove Singularity} \label{sec:vhs}

This is a RPA spin excitation theory for the recently reported INS
data on YBCO. For simplicity, we have mimic the realistic
$extended$-saddle-vHS at the proximity of Fermi level
\cite{GCA94,GCD93} by simple saddle-vHS just beneath Fermi level.
It is believed that in RPA theories they will be qualitatively the
same. Nevertheless its extended nature is crucial for bare theory
to have quantitative agreement with experiments \cite{Abr00}.

%%% chit-chat

A number of comprehensive theories including both commensurate and
IC excitation \cite{KSL00,Nor00,Abr00,VW99,OP99,BL99} were
explored since the discovery of the behavior in YBCO. It is
important to note that all of them are Fermi-liquid-like theories,
and most are RPA theories~\cite{KSL00,Nor00,OP99,BL99}. The
possibility of a bare theory was investigated by Abrikosov
\cite{Abr00,Abr98} and us \cite{VW99}. The commensurate peaks in
bare theories are generally weaker, broader, and occurs at
$\omega_o\gtrsim2\Delta$. That is in contrast to the much stronger
and sharper one in $\chi''$ that occurs at $\omega_o<2\Delta$.
However, recently it has been shown that incorporating the
extended-vHS \cite{Abr00} into bare theory can also give peak
intensity and width in reasonable agreement with experiments.
Attempting to rely on the criterion $\omega_o$ being less or
greater than $2\Delta$ for choosing a theory also encounters the
subtleties in the determination of the SC gap. Especially in the
underdoped samples where there are still controversies on whether
the SC gap and the pseudogap are the same gap
\cite{KYW00,SYN99,GVM99}. Therefore either
$\omega_o\gtrsim2\Delta$ (see discussion in Ref.~\cite{Abr98}) or
$\omega_o<2\Delta$ (e.g. in the underdoped compounds) can be
occasionally favored. As a result we believe that at present there
is no clear evidence here to favor either scenarios. Recently the
IC peaks have received more attention because of their
frequency-shifted behavior. Most of the present scenarios for it
are based on the nesting effect which depends mainly on the nodal
FS, except one recently proposed by Abrikosov \cite{Abr00} depends
on the band mass at the extended-vHS region. The discussion of the
temperature-shifted IC peaks has been missing from the literature.

%%% beginning

With a sufficiently strong correction strength, it is always
possible to have a RPA resonance at ${\bf Q}_{AF}$. The reason is
since any susceptibility function or interaction vertex on the
lattice is symmetric at ${\bf Q}_{AF}+\delta{\bf q}\rightarrow{\bf
Q}_{AF}-\delta{\bf q}$, they always have an extremum at ${\bf
Q}_{AF}$ [e.g. see Fig.~\ref{rpa2}] and thus one can always have
$1-V\chi_o'=0$ locally at ${\bf Q}_{AF}$.

%%% zero temperature spectrum
\begin{figure}
\begin{center}
\vspace{-1.0cm}
\leavevmode\epsfxsize=3.4in \epsfbox{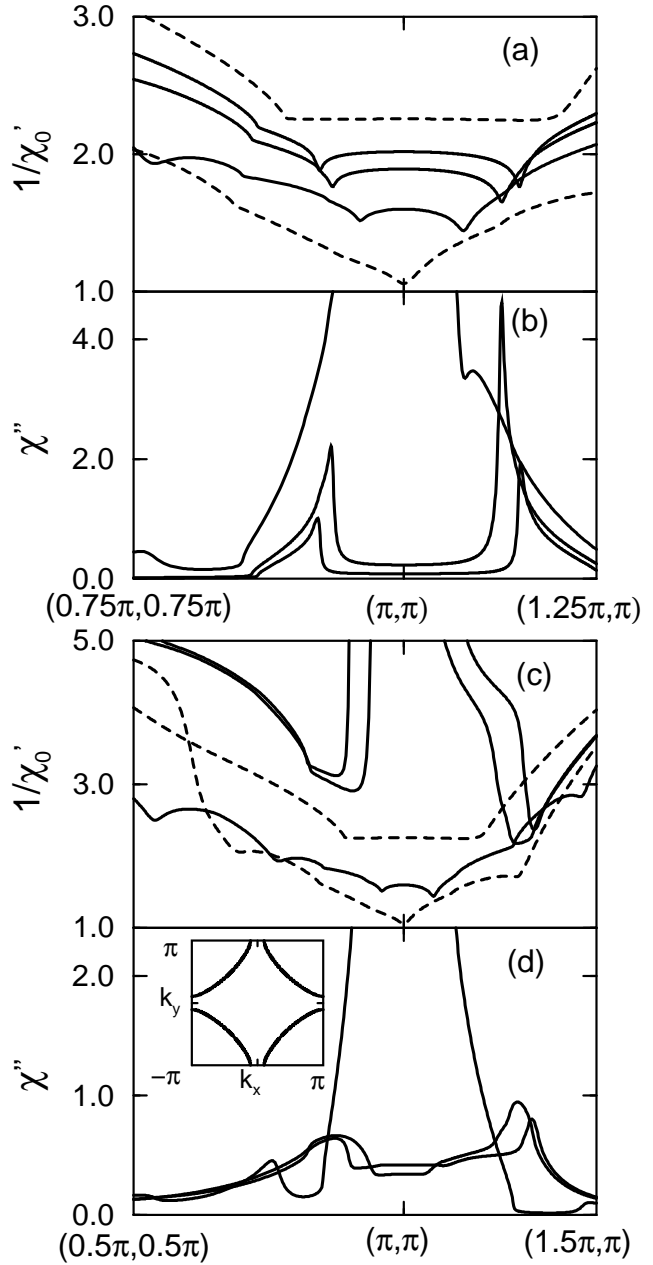}
\vspace{-0.0cm}
\end{center}
\caption{Frequency-driven peak shifting at different frequency
regimes : (a) $1/\chi_o'({\bf q},\omega)$ at
$\omega/\Delta=1.0,1.2$, and $1.5$ [solid lines from top to
bottom]; (b) $\chi''({\bf q},\omega)$ at $\omega/\Delta=1.0,1.2$,
and $1.5$ [at decreasing incommensurability along
$(\pi,\pi)-(1.25\pi,\pi)$]. (c) $1/\chi_o'({\bf q},\omega)$ at
$\omega/\Delta=1.5,2.3$, and $2.5$ [solid lines at increasing
``incommensurability'' of the minima along
$(\pi,\pi)-(1.5\pi,\pi)$]; (d) $\chi''({\bf q},\omega)$ at
$\omega/\Delta=1.5,2.3$, and $2.5$ [at increasing
incommensurability along $(\pi,\pi)-(1.5\pi,\pi)$]. The dashed
lines from top to bottom in (a) and (c) show $1/\chi_o'$ at
$\omega/\Delta=0$ and $1.77$ respectively. $\chi''$ in (b) and (d)
are calculated with $|J|/t=1.6$, and the commensurate resonance
occurs at $\omega=\omega_o(T=0)=1.5\Delta(0)$. We have truncated
the strong commensurate peak for obvious reason. $T=0$, the
quasiparticle dispersion is $t=1$, $t'=-0.20$, $\mu=-0.65$, and
$\Delta=0.30$ (gives $\xi_{{\bf k}={\overline M}}/\Delta=-0.5$).
The inset shows the FS. The broadening taken for all the
calculations in this section is $\Gamma/t=0.008$.}
\label{vhs1} \vspace{-0.3cm}
\end{figure}

% x= -0.35 to 0.25, (a) y= 1.0 to 3.0 and (b) 0 to 4.8;
% x= -0.7 to 0.5, (c) y= 1.0 to 5.0 and (d) 0 to 2.4 (which is 4.8/2).
%%%
The dispersion taken here to mimic the YBCO dispersion has a
hole-like FS and a simple saddle-vHS at ${\overline M}=(0,\pi)$
slightly beneath the Fermi level ($\xi_{{\bf k}={\overline
M}}/\Delta=-0.5$). The interaction is chosen as an
antiferromagnetic $J_{\bf q}$ with $|J|/t=1.6$. At $T=0$, a
commensurate resonance occurs at $\omega/\Delta=1.5$. The momentum
dependence of the RPA spectrum and $1/\chi_o'$ are shown in
Fig.~\ref{vhs1}, at different frequencies and zero temperature.
The onset of $\chi_o''({\bf Q}_{AF},\omega)$ and the minimum of
$1/\chi_o'({\bf Q}_{AF},\omega)$ occurs at $\omega/\Delta=1.77$.

%%%
\begin{figure}[t]
\begin{center}
\vspace{-0.0cm} \leavevmode\epsfxsize=3.3in \epsfbox{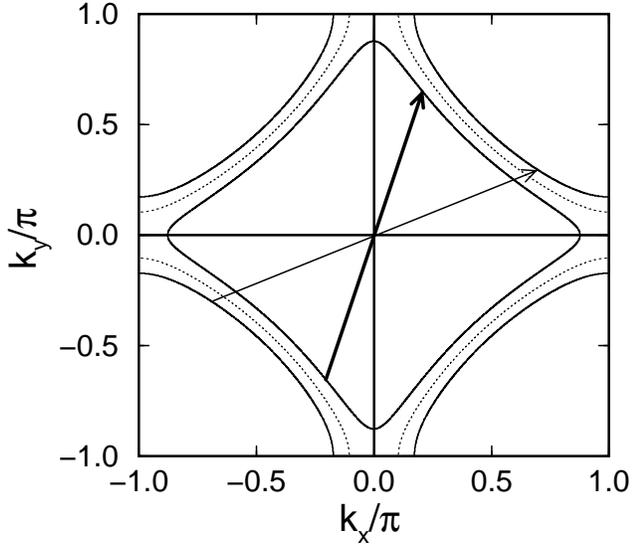}
\vspace{-0.3cm}
\end{center}
\caption{Energy contour $E_{\bf k}=\omega/2$ for
$\omega=2.5\Delta(0)>\omega_o(T=0)$. The thick local-nesting
vector between $below$-FS contour has more pronounced nesting
effect (compared to the thin vector) due to the smaller curvature.
We remind that the dispersion here has a saddle-vHS near to Fermi
level (refer the exact parameters to Fig.~\protect\ref{vhs1}). The
FS is indicated as dotted line.}
\label{vhs2} \vspace{-0.3cm}
\end{figure}
%%%
A new feature appears at $\omega>\omega_o(T=0)$ due to the shallow
saddle-vHS. The discommensuration is seen to increase at
increasing frequency [see Fig.~\ref{vhs1}(d)]. It is also due to a
dynamic local nesting effect similar to that at
$\omega\rightarrow\omega_o$ from below. Owing to the shallow flat
band at ${\overline M}$, the nesting contours $E_{\bf k}=\omega/2$
at high frequencies necessarily have smaller curvature at those
segments below FS, and can be more effectively nested locally [see
Fig.~\ref{vhs2}]. The peak profile here is weak because the below-
and above-FS contours are too far apart to coherently form a
ridge. The RPA correction has an effect similar to that at
$\omega<\omega_o(T=0)$.

%%%
Replacing the saddle-vHS by an extended-saddle-vHS do not affect
the IC excitation at $\omega<\omega_o$ since only the nodal FS is
of concern. Neither the diverging discommensuration at
$\omega>\omega_o$ will be affected since the switch-over of the
smaller and larger curvature energy contours should also occur.

%%%
The large change in $1/\chi_o'$ at $\omega/\Delta=0\sim2$ is a
result of the large jump at the onset of $\chi_o''({\bf
Q}_{AF},\omega)$ (at
$\omega_{\delta'}/\Delta(T)=\sqrt{\mu/t'}\simeq1.8$). The onset is
sharp and large due to the enhancement by the spin coherence
factor (at $\Delta_{\bf k}\Delta_{{\bf k}+{\bf Q}_{AF}}<0$), and
the picking up of states around the shallow vHS at ${\overline M}$
(at a depth of $\xi_{{\bf k}={\overline M}}=4t'-\mu$) by the
$\delta$-function [see Eq.~(\ref{eq:chi011t0})]. This widens the
window and lowers the threshold for the interaction strength
giving resonance. This may partially account for the existence of
the commensurate peak in YBCO and BSCCO, but not in LSCO. The RPA
effect at ${\bf Q}_{AF}$ is effectively turned off when $\omega$
is increased beyond the maximum of $\chi_o''({\bf
Q}_{AF},\omega)$, since $1/\chi_o'({\bf Q}_{AF},\omega)$ recedes
rapidly from zero then [see Fig.~\ref{vhs1}(c)]. Since the onset
frequency scales with $\Delta(T)$, the RPA commensurate resonance
is necessarily softened by temperature. But the softening could be
very little. The reason is that as $\Delta(T)$ gets smaller,
divergence in $\chi_o'({\bf Q}_{AF},\omega)$ gets smaller and will
fail to fulfill the resonance condition $\chi_o'({\bf
Q}_{AF},\omega)=1/V({\bf Q}_{AF})$ for a given $V({\bf Q}_{AF})$.
The peak could cease to exist before its softening is really
detected. The case for the softening of the commensurate peak in a
bare theory is similar \cite{VW99}.

%%% finite temperature
\begin{figure}[t]
\begin{center}
\vspace{-0.0cm} \leavevmode\epsfxsize=3.2in \epsfbox{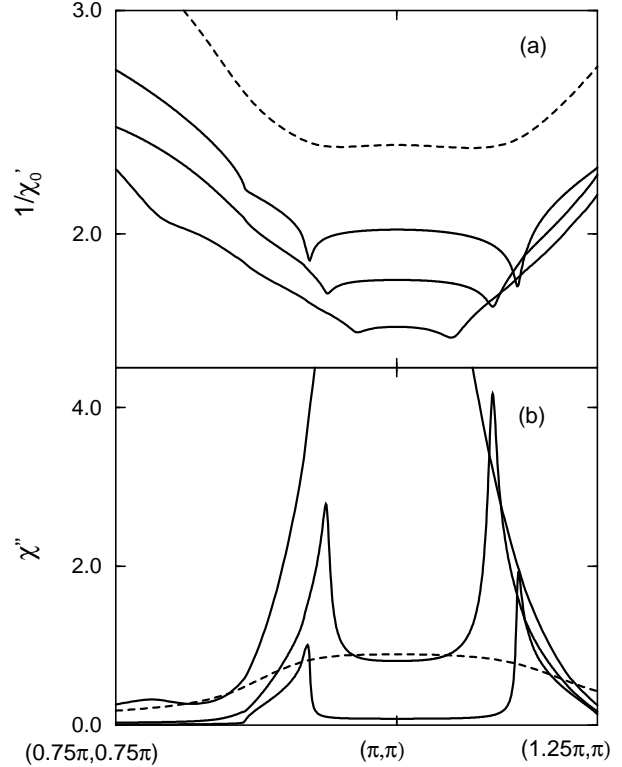}
\vspace{-0.0cm}
\end{center}
\caption{Temperature-driven peak shifting at low frequency
regime~: At $\omega=1.0\Delta(0)<\omega_o(T=0)$, (a)
$1/\chi_o'({\bf q},\omega)$ is shown at $T/T_c=0,0.8,0.88$ (solid
lines from top to bottom), and $1.0$ (dashed line); (b)
$\chi''({\bf q},\omega)$ is shown at $T/T_c=0,0.8,0.88$ (solid
lines at decreasing incommensurability), and $1.0$ (dashed line).
The weak commensurate structure in the normal state is seen to
grow into a strong peak at $T=0.88T_c\protect\lesssim T_c$, then
splits into IC peaks which recede from ${\bf Q}_{AF}$ at
$T\rightarrow0$. The strong commensurate peak is truncated for
obvious reason. $T_c=0.25\Delta(0)$, and the $T$-dependence of
$\Delta(T)$ is assumed as the same as in
Fig.~\protect\ref{surc1}(b). Other parameters are referred to
Fig.~\protect\ref{vhs1}.}
\label{vhs3} \vspace{-0.3cm}
\end{figure}

% x= -0.35 to 0.25, (a) y=1.4 to 3.0; (b) y= 0 to 4.5

%%%
Figure~\ref{vhs3} plots the temperature-evolution of a low
frequency spectrum in momentum space. The most notable should be
the softened commensurate resonance at
$\omega\lesssim\omega_o(T=0)$. When the system is cooled down from
above $T_c$, the broad and weak commensurate peak in the normal
state grows into a sharp and strong peak at some temperature just
below $T_c$, then it is split up into IC peaks and recede from
${\bf Q}_{AF}$, to a fixed location at low temperature. The weak
IC structure at $\omega>\omega_o(T=0)$ in SC state is found to be
simply smeared off to a broad commensurate at warming up to the
normal state \cite{VCW00}.

%%%
Note that the observed relative excitation intensity in the course
of the evolution [see Fig.4 in Ref.~\cite{BSF00}] can only be
obtained correctly in $\chi''({\bf q},\omega)$ via an appreciable
interaction strength. Though the ``shoot and split'' behavior of
the $\omega<\omega_o(T=0)$ spectrum at cooling down is already
contained in a bare theory invoking vHS effect for the
commensurate peak \cite{VW99}, the normal state commensurate
structure always incorrectly has an intensity higher than the low
temperature IC structure in $\chi_o''({\bf q},\omega)$ [see, for
example, a similar case in Fig.~\ref{surc1}(b)].

%%%
\begin{figure}[t]
\begin{center}
\vspace{-0.0cm} \leavevmode\epsfxsize=3.2in \epsfbox{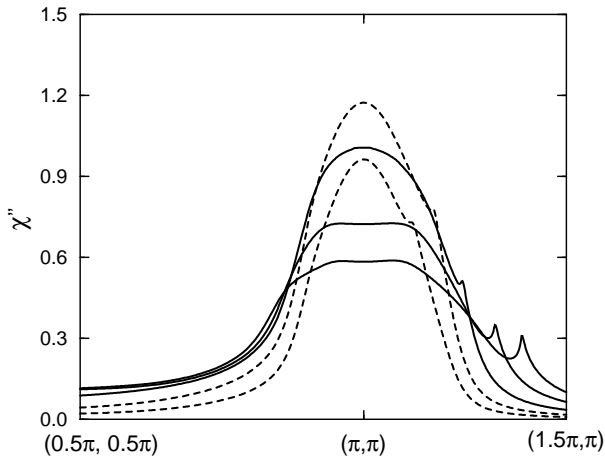}
\vspace{+0.3cm}
\end{center}
\caption{The broad commensurate peak of normal
state $\chi''({\bf q},\omega)$ at $\omega/\Delta(0)=0.2,0.4$
(dashed lines from bottom to top), $0.8,1.4$, and $2.0$ (solid
lines from top to bottom). $T=T_c=0.25\Delta(0)$. Refer those not
stated parameters to Fig.~\protect\ref{vhs1}.}
\label{vhs4} \vspace{-0.3cm}
\end{figure}
%%%
A weak and broad commensurate structure is always seen in the
normal state at all frequencies [see Fig.~\ref{vhs4}] as what is
observed in experiments \cite{BSF00,ANE99}. It is due to the
enhancement by the saddle-vHS and RPA correction. For the model
system considered here, its intensity reaches a maximum at some
$\omega/\Delta(0)\simeq0.4$ as a result of the competition between
$\chi_o''$ and RPA enhancement.

%%% ending

In summary, the diverging incommensurability at $\omega>\omega_o$
in the SC state is accounted by the approaching of ${\overline
M}$-point vHS to the Fermi level; the normal state broad
commensurate structure at any frequency also finds a natural
explanation in it. The recent observation on the relative
excitation intensity in the normal and SC state prefers a RPA
theory in the resonance regime more than the bare theories, though
the gross features of the bare and RPA spectra are the similar.
Including an underdoped sample (by Arai $et$ $al.$ \cite{ANE99})
and a nearly optimally doped sample (by Bourges $et$ $al.$
\cite{BSF00}), the converging and diverging shifting of the IC
peaks has a transparent explanation in the FL picture.

\section{anisotropy effect}
\label{sec:aniso}

Anisotropy effect in the high-$T_c$ cuprates was discussed before
by Rendell and Carbotte \cite{RC96}. The effect is believed to be
important in systems such as the orthorhombic crystals or the
chained-YBCO. In those systems, anisotropy may either exist in the
band dispersion or enter into the SC order parameter, or exist in
both. It was pointed out in Ref.~\cite{RC96} that if only one of
those two kinds of anisotropy exist in the system, it can be
distinguished by the criterion that the gap anisotropy exist only
in the SC state while the dispersion anisotropy persists into the
normal state. Some important effects of it which can be seen in
INS spectrum are, if there is an admixture of $s$-component in the
$d$-wave gap then there should be a rendering of the four-fold
symmetry of the quasielastic node-to-node IC peaks to two-fold,
and regardless of the anisotropy source a difference of the
nesting IC peak intensities on different crystal axis should be
observed.

%%%
We will discuss here a case of anisotropy in the gap, the
existence of an $s$-component and the SC order parameter has the
form $\Delta_{\bf k}=\Delta(T)[a s_{\bf k}+(1-a) d_{\bf k}]$,
where $s_{\bf k}=1$ and $d_{\bf k}=(\cos k_{x}-\cos k_{y})/2$. The
existence of an $s$-component is suggested by the tunneling
experiments \cite{KSC97,KKS96,SGM94} and its percentage is at
$a\sim0.2$. It was discussed in Ref.~\cite{RC96} at low
frequencies, but in order to clearly distinguish the effect we
think that it is necessary to study its $\omega$-dependence.
%%%
\begin{figure}[h]
\begin{center}
\vspace{-0.0cm} \leavevmode\epsfxsize=3.4in \epsfbox{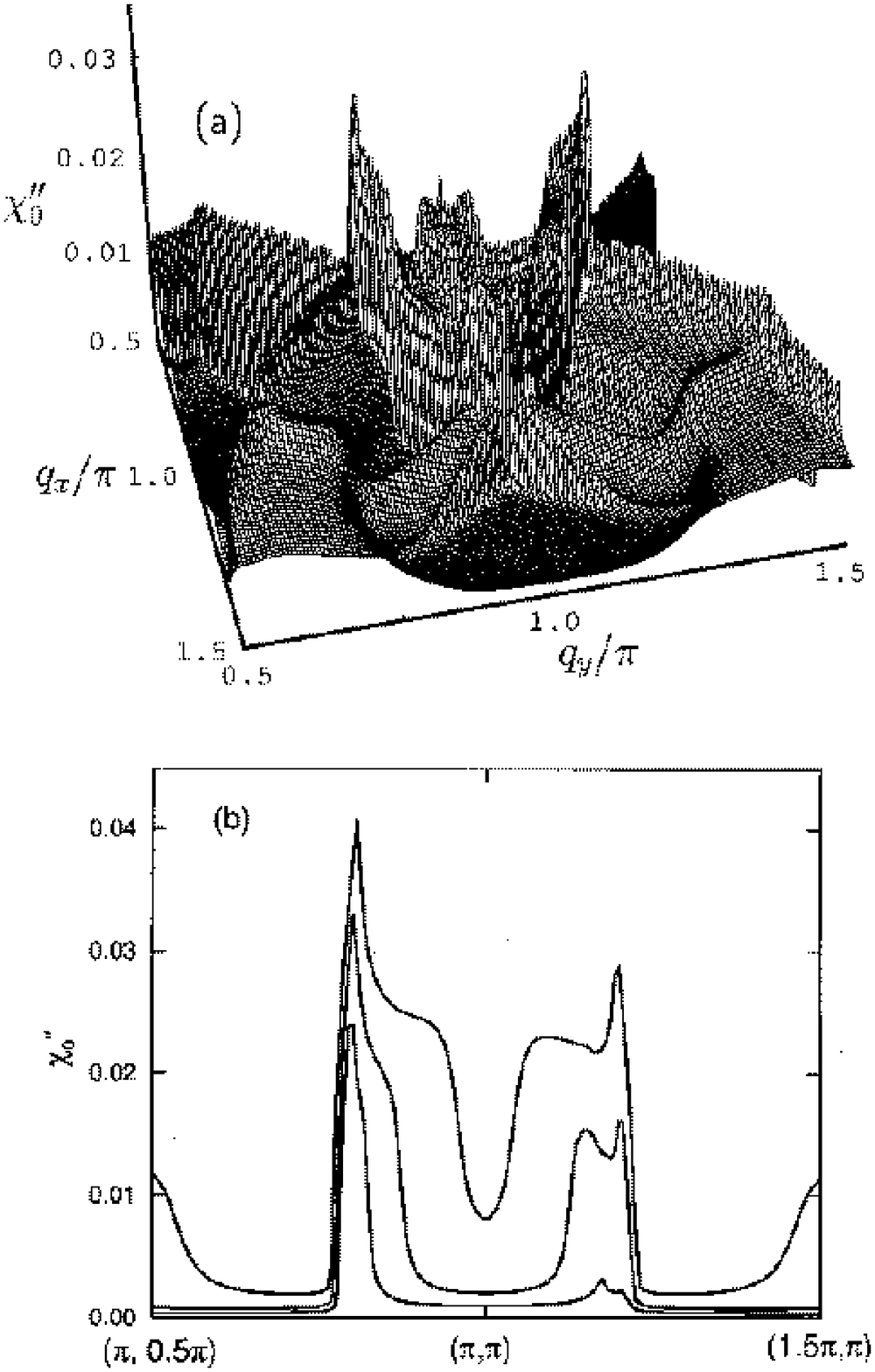}
%\vspace{-1.5cm}
\end{center}
\caption{Anisotropic SC state $\chi_o''({\bf q}, \omega)$ (a) in
the full {\bf q}-space at $\omega/\Delta=1.0$, and (b) along ${\bf
q}=(\pi,0.5\pi)-(\pi,\pi)-(1.5\pi,\pi)$ at
$\omega/\Delta=0.6,0.9$, and $1.2$ (from bottom to top). The
difference in height of the peaks on different axis is more
spectacular at low frequencies. The SC order parameter is taken as
$\Delta_{\bf k}=\Delta[0.2s_{\bf k}+0.8d_{\bf k}]$, where
$\Delta=0.10$, $s_{\bf k}=1$, and $d_{\bf k}=(\cos k_{x}-\cos
k_{y})/2$. $T=0$ and the dispersion is $t=1$, $t'=-0.25$, and
$\mu=-0.65$.}
\label{aniso1} \vspace{-0.3cm}
\end{figure}
%%%

Figure~\ref{aniso1} shows the contrast of the nesting IC peaks
along different crystal axis at different $\omega$, on a system
with reasonable size of $s$-component in the gap. The difference
in peak intensities is seen to be most prominent at low $\omega$,
and rapidly decreased at increasing frequency. Since the data by
Mook $et$ $al.$ (on a detwinned orthorhombic
YBa$_2$Cu$_3$O$_{6.6}$) \cite{MDD00} is at a somewhat low
frequency $24$ meV $\sim\Delta(0)$, we believe data on a wider
$\omega$ range is desirable in order to make definite statement on
the origin of the one-dimensional nature.

%%%
The difference in peak intensities can be easily explained by the
picture given in Sec.~\ref{sec:dln}. The twofold symmetric
spectrum has the node-to-node excitation closer to one of the two
crystal axis, and excitation away from the nodal excitation is
suppressed by the coherence factor, therefore the peaks on one of
the axis has lower intensity. This can only dominate at
frequencies as low as $\sim\Delta$ and excitation along the FS has
not been fully opened up.

%%%
We have presented here an anisotropy effect suggested by the
tunneling experiment and it can introduce one-dimensional feature
into the INS spectrum (other possibility is also suggested in
Ref.~\cite{BKR00}). The frequency dependence of the contrast of IC
peak intensity should act as a criterion to compare with the
Stripe interpretation which has no frequency dependence.
Furthermore such gap anisotropy effect should vanish in the normal
state.

\section{discussion and conclusions}
\label{sec:conc}

We have presented a thorough study on the basic properties of the
spin susceptibility and have summarized our findings and
comparisons with experiments at the end of each section.

%%%
Some possible departure of our conclusions from reality should be
mentioned. The presence of impurity in real systems destroys the
exact translational symmetry and mixes states of different
momentum, which are also the energy eigenstates. Such process
necessarily relaxes the energy conservation selection of the
transition region in phase space. The effect should be significant
in low energy processes in the SC state where there is
inhomogeneous gapping out of phase space by the $d$-wave gap. For
example, it may pull down the lower threshold for the existence of
the nesing IC peaks \cite{QS95}. We have also taken a simple
$d$-wave in the discussion which is just meant to describe a
symmetry with four nodes and change of sign. The actual {\bf
k}-dependence of the gap should exhibit some deviation from pure
$d$-wave since other factors such as interlayer tunneling
\cite{YCA97} or correlation effects $etc.$ may influence. The
important case of possessing a $s$-component was discussed in
Sec.~\ref{sec:aniso}. The realistic dispersion at the FS also
shows deviation from our simple tight-binding band, the most well
known of all should be the presence of extended-vHS. Therefore all
numerals we give are qualitative.

%%%
To a qualitative level, the ``inconsistency'' between INS spectra
of LSCO and YBCO as mentioned in Sec.~\ref{sec:intro} can find a
consistent picture in the FL interpretation. They could be systems
in different regimes of the ratio $\Delta/v_F$. The observations
on LSCO [as discussed in Sec.~\ref{sec:survey} and
Sec.~\ref{sec:rpa}]: the wide existing IC excitation with
incommensurability depends only on doping, can be ascribed to a
small $\Delta/v_F$ of the system. The observations on YBCO [as
discussed in Sec.~\ref{sec:survey} and Sec.~\ref{sec:vhs}]:
including the $\omega$- and $T$-driven shifted IC peaks in SC
state, and normal state broad commensurate peak, can be all
ascribed to an appreciable $\Delta/v_F$ of the system. The
diverging incommensurability at $\omega>\omega_o$ in SC state are
ascribed to the approaching of the ${\overline M}$-point vHS to
the Fermi level. Though bare and RPA theories could be
qualitatively similar, RPA theories can describe the temperature
evolution of the excitation intensity more adequately. A peculiar
but straightforward conclusion for the commensurate resonance is
it can occur at $\omega\lesssim\omega_o(T=0)$ when $T\lesssim
T_c$, either in the bare \cite{VW99} or RPA theories. This
naturally explains the observed enhancement of scattering
intensity at ${\bf Q}_{AF}$ in that regime [see Fig.4(c) in
Ref.~\cite{BSF00}].

%%%
We would like to point out that while the observed commensurate
peak in the SC state could easily find an interpretation as a
resonance, as it occurs at a small frequency window; the lower
intensity broad commensurate peak in the normal state which is
observed at $any$ frequency is hard to be interpreted in terms of
a resonance. In our scenario this is a natural consequence of the
overlap of broad IC peaks, with further enhancement by the ${\overline
M}$-point vHS and RPA-correction.

%%%
On the ratio $\Delta/v_F$ of LSCO and YBCO, there should be a
consensus that the SC gap of LSCO ($\sim$10 meV) is a fraction of
that of YBCO ($\sim$30 meV). Due to the low resolution in the
determination of electronic dispersion (the resolution in ARPES
measurements is about 10-40 meV), presently $v_F$ in most material
could not be precisely determined \cite{note1,note2,BLK00} and
neither the ratio $\Delta/v_F$. Therefore in this paper we have
meant only a first-step qualitative study, and did not predict any
precise figure for $v_F$ of different systems. But it is still
important to note that in our scenario, $v_F$ of YBCO should be at
least several times smaller than that of LSCO .

%%%
In the literature, there are also other approaches \cite{KSL00,Nor00,Abr00,CJT00} to
the issue. As we do, Ref.~\cite{KSL00} also simultaneously
accounts for both the LSCO and YBCO data, but does not elaborate
on how the difference in the behavior of the IC peak arises.
Furthermore, it is predicted that a commensurate peak should also
exist in LSCO. Ref.~\cite{Nor00} addresses only the YBCO data by
using a refined tight-binding band. It is also not shown
explicitly that how the IC peaks are shifted. Moreover, the
relation to the LSCO data is not mentioned. Ref.~\cite{Abr00}
obtains the commensurate and shifting IC peaks from the
quasiparticle masses at the extended-saddle-vHS region, while the
data of LSCO is also not mentioned. In our present work, we assume
that the IC peaks in LSCO and YBCO have the same origin, and have
related the different behaviors of them to their individual Fermi
velocities $v_F$. An explicit explanation of the shifting behavior
is also given. The approach in Ref.~\cite{CJT00} is very different
from ours as it interprets the shifting of the IC peak as the
downward dispersion of a spin wave mode. Our approach can also
easily address the shifting of IC peak caused by temperature
change, which is so far only discussed by us.

%%%
There are some bewildering stems from our work. In the underdoped
cuprates, anomalies such as the pseudogap phenomenon or the
destruction of FS \cite{MDL96} are seen near the maximum gap
region and it is likely that the FL picture is to breakdown at
that region. But now the FL behavior near the maximum gap region
is an important ingredient in our scenario on the YBCO, and our
coverage includes both the underdoped and optimally doped YBCO. It
seems that to some extent the FL behavior is still retained in the
response to INS measurements.

%%%
In conclusion, albeit the FL picture on the cuprates is recently
quested by many experiments, most behaviors of the recent INS data
are still well described in terms of the traditional Fermi liquid
picture.

\acknowledgments KKV acknowledges the scholarship from NSC of
Taiwan under grant No.89-2112-M-003-009. We thank P. Bourges $et$
$al.$ for sending us their results prior to publication. We also
want to thank A. Abrikosov for sending us his preprint as well as
useful correspondences.


\begin{references}
\bibitem{note2}
We want to emphasize that the dispersion is indeed badly known because of the
  low resolution of ARPES measurements. Even in the long studied compound
  BSCCO, it was until recently that some kink in the dispersion is realised.
  The kink results in a reduction of velocity by a factor of 2 or more. It is
  interesting to note that BSCCO is a compound structurally similar to YBCO,
  and we actually need a $v_F$ of YBCO several times smaller than that of LSCO.
\end{references}
\end{document}